\documentclass[fleqn,10pt]{achemso}
\usepackage[utf8]{inputenc}
\usepackage[T1]{fontenc}
\usepackage{algorithm}
\usepackage[braket, qm]{qcircuit}
\usepackage{graphicx}
\usepackage{hyperref} 
\usepackage{physics}
\usepackage{acro}
\usepackage{subcaption}
\usepackage{algpseudocode} 
\usepackage{siunitx}
\usepackage[section]{placeins}
\usepackage{rotating}
\DeclareMathOperator*{\argmax}{arg\,max}

\DeclareAcronym{hf}{
    short = HF ,
    long = Hartree-Fock,
}
\DeclareAcronym{dft}{
    short = DFT,
    long = Density Functional Theory,
}
\DeclareAcronym{qm}{
    short = QM,
    long = quantum mechanics,
}
\DeclareAcronym{qc}{
    short = QC,
    long = quantum computing,
}
\DeclareAcronym{wfdft}{
    short = wf@DFT,
    long = wave function in DFT projective embedding,
}
\DeclareAcronym{mm}{
    short = MM,
    long = molecular mechanics,
}
\DeclareAcronym{qmmm}{
    short = QM/MM,
    long = quantum mechanics/molecular mechanics,
}
\DeclareAcronym{xtb}{
    short = xTB,
    long = extended tight-binding,
}

\DeclareAcronym{neb}{
    short = NEB,
    long = nudged elastic band,
}
\DeclareAcronym{vqe}{
    short = VQE ,
    long = variational quantum eigensolver,
}
\DeclareAcronym{adapt}{
    short = ADAPT-VQE ,
    long =  Adaptive Derivative-Assembled Pseudo-Trotter Ansatz Variational Quantum Eigensolver,
}
\DeclareAcronym{fast}{
    short =  FAST-VQE,
    long = Fermionic Adaptive Sampling Theory VQE,
}
\DeclareAcronym{nisq}{
    short = NISQ ,
    long = Noisy Intermediate-Scale Quantum,
}
\DeclareAcronym{jw}{
    short = JW ,
    long = Jordan-Wigner,
}
\DeclareAcronym{dmet}{
    short = DMET,
    long = density matrix embedding theory,
}
\DeclareAcronym{gfnff}{
    short = GFN-FF,
    long = {Geometry, Frequency, Noncovalent-Force Field},
}
\DeclareAcronym{gfn2xtb}{
    short = GFN2-xTB,
    long = {Geometry, Frequency, Noncovalent, extended Tight Binding v2},
}
\DeclareAcronym{casci}{
    short = CASCI,
    long = {complete active space configuration interaction},
}
\DeclareAcronym{fci}{
    short = FCI,
    long = {full configuration interaction},
}
\DeclareAcronym{ca}{
    short = CA,
    long = carbonic anhydrase,
}
\DeclareAcronym{alpb}{
    short = ALPB,
    long = analytical linearized Poisson–Boltzmann,
}
\DeclareAcronym{qeb}{
    short = QEB,
    long = qubit excitation-based,
}

\DeclareSIUnit\angstrom{\text {Å}}
\DeclareSIUnit\atomicunit{a.u.}
\DeclareSIUnit\calorie{cal}

\title{Calculating the energy profile of an enzymatic reaction on a quantum computer.}

\author{Patrick Ettenhuber}
\email{pe@kvantify.dk}
\affiliation{Kvantify Aps, DK-2100 Copenhagen, Denmark}
\author{Mads B{\o}ttger Hansen}
\affiliation{Kvantify Aps, DK-2100 Copenhagen, Denmark}
\author{Pier Paolo Poier}
\affiliation{Kvantify Aps, DK-2100 Copenhagen, Denmark}
\author{Irfansha Shaik}
\affiliation{Kvantify Aps, DK-2100 Copenhagen, Denmark}
\alsoaffiliation{Department of Computer Science, Aarhus University, DK-8000 Aarhus C, Denmark}
\author{Stig Elkjær Rasmussen}
\affiliation{Kvantify Aps, DK-2100 Copenhagen, Denmark}
\alsoaffiliation{Department of Physics and Astronomy, Aarhus University, DK-8000 Aarhus C, Denmark}
\author{Niels Kristian Madsen}
\affiliation{Kvantify Aps, DK-2100 Copenhagen, Denmark}
\author{Marco Majland}
\affiliation{Kvantify Aps, DK-2100 Copenhagen, Denmark}
\alsoaffiliation{Department of Physics and Astronomy, Aarhus University, DK-8000 Aarhus C, Denmark}
\alsoaffiliation{Department of Chemistry, Aarhus University, DK-8000 Aarhus C, Denmark}
\affiliation{Kvantify Aps, DK-2100 Copenhagen, Denmark}
\author{Frank Jensen}
\alsoaffiliation{Kvantify Aps, DK-2100 Copenhagen, Denmark}
\affiliation{Department of Chemistry, Aarhus University, DK-8000 Aarhus C, Denmark}
\author{Lars Olsen}
\affiliation{Novonesis A/S, DK-2800 Kgs. Lyngby, Denmark}
\author{Nikolaj Thomas Zinner}
\alsoaffiliation{Department of Physics and Astronomy, Aarhus University, DK-8000 Aarhus C, Denmark}
\affiliation{Kvantify Aps, DK-2100 Copenhagen, Denmark}

\AtBeginDocument{\RenewCommandCopy\qty\SI}
\begin{document}
\begin{abstract}
Quantum computing (QC) provides a promising avenue toward enabling quantum chemistry calculations, which are classically impossible due to a computational complexity that increases exponentially with system size. As fully fault-tolerant algorithms and hardware, for which an exponential speedup is predicted, are currently out of reach, recent research efforts are dedicated to developing and scaling algorithms for \ac{nisq} devices to showcase the practical usefulness of such machines. To demonstrate the usefulness of NISQ devices in the field of chemistry, we apply our recently developed FAST-VQE algorithm and a state-of-the-art quantum gate reduction strategy based on propositional satisfiability together with standard optimization tools for the simulation of the rate-determining proton transfer step for $\text{CO}_2$ hydration catalysed by carbonic anhydrase resulting in the first application of a quantum computing device for the simulation of an enzymatic reaction. To this end, we have combined classical force field simulations with quantum mechanical methods on classical and quantum computers in a hybrid calculation approach. The presented technique significantly enhances the accuracy and capabilities of QC-based molecular modeling and finally pushes it into compelling and realistic applications. The framework is general and can be applied beyond the case of computational enzymology. 
\end{abstract}
\section{Introduction}
Enzymes are remarkable biological molecules that act as catalysts. They can speed up chemical reactions, often by many orders of magnitude, under the moderate conditions that living organisms thrive in. At the same time, enzymes can be exceptionally selective towards target substrates, making chemical reactions extremely fast while reducing undesirable side reactions\cite{voet2016, metzler2001, Hedstrom_2010, Shaik2010}.\\
These striking catalytic properties are the result of natural evolution that has slowly optimized and shaped enzymes as we know them today. Understanding the underlying catalytic mechanism is not only interesting in its own right, but it also holds immense potential to improve our quality of life. For example, this knowledge can pave the way for developing novel life-saving drugs\cite{Williams2017}, cleaner bioremediation procedures\cite{Mousavi2021}, and more sustainable chemical processes\cite{Reisenbauer2024}. In medicinal chemistry, for example, the inhibition of enzymes via selective inhibitors can have many therapeutic applications, a prime example is antibiotics that kill harmful bacteria in a human host by selectively blocking enzymes in central bacterial metabolic pathways\cite{metzler2001, voet2016}. From an industrial point of view, the replacement of traditional catalysts with enzymes could make large-scale processes more sustainable by reducing operating temperatures and pressures and by replacing organic solvents with water\cite{Reisenbauer2024}. Thus, considerable effort has been spent elucidating the underlying principles of enzymatic catalysis, both experimentally and computationally. In fact, computational enzymology has become an important tool that complements experimental studies\cite{Warshel_1972, Levitt_1975, Warshel_1976, Senn2007, Shaik2010} of mechanisms and it can be divided into different topics, each characterized by dedicated computational approaches\cite{Senn2007, Shaik2010, Gerlt2017, Culca2017}. \\
The large number of calculations involved in ligand screening for drug discovery often demands the use of parameterized energy functions which are computationally cheap but not suitable for calculating chemical properties. Such parametrized energy functions are typically not applicable for calculating kinetic constants derived from energy barriers, as encountered in enzymatic catalysis. For this purpose, quantum mechanical methodologies are required and it is generally attempted to obtain chemical accuracy, defined as 1 kcal/mol, for quantitative predictions\cite{Pople1998}. Kohn-Sham \ac{dft} is the most widely used electronic structure method in computational chemistry as it introduces electronic correlation at a moderate computational cost\cite{jensen2016}. While in principle mathematically exact, the true form of the exchange-correlation functional is unknown but a range of approximate functionals with varying accuracies is available\cite{jensen2016, Mardirossian2017}. Furthermore, the lack of a clear indicator for reliable results can make it difficult to assess the actual accuracy of DFT results, diminishing its value as a predictive tool.
Wave function-based correlated methods, on the other hand, offer the potential for exact solutions\cite{olsen_bible}. However, this comes at a significant computational cost making it only practical for small molecules. \\
Despite tremendous progress wave function-based methods and DFT are not applicable for full proteins. Catalytic mechanisms, however, usually occur in small regions of enzymes, making it thus possible to separate the systems into regions of low and high accuracy. These multi-scale modeling methods have already been developed in the '70s by Levitt and Warshel\cite{Warshel_1972, Levitt_1975, Warshel_1976}. Wave function methods or, more often, DFT may be used for the high-accuracy calculations\cite{kalek_2017, Calcagno2024}, but since these multiscale methods require rather large high-accuracy regions the disadvantages of DFT and wave function-based methods still apply and therefore novel approaches are desired. \\

\begin{figure}
    \centering
    \includegraphics[scale=1.3]{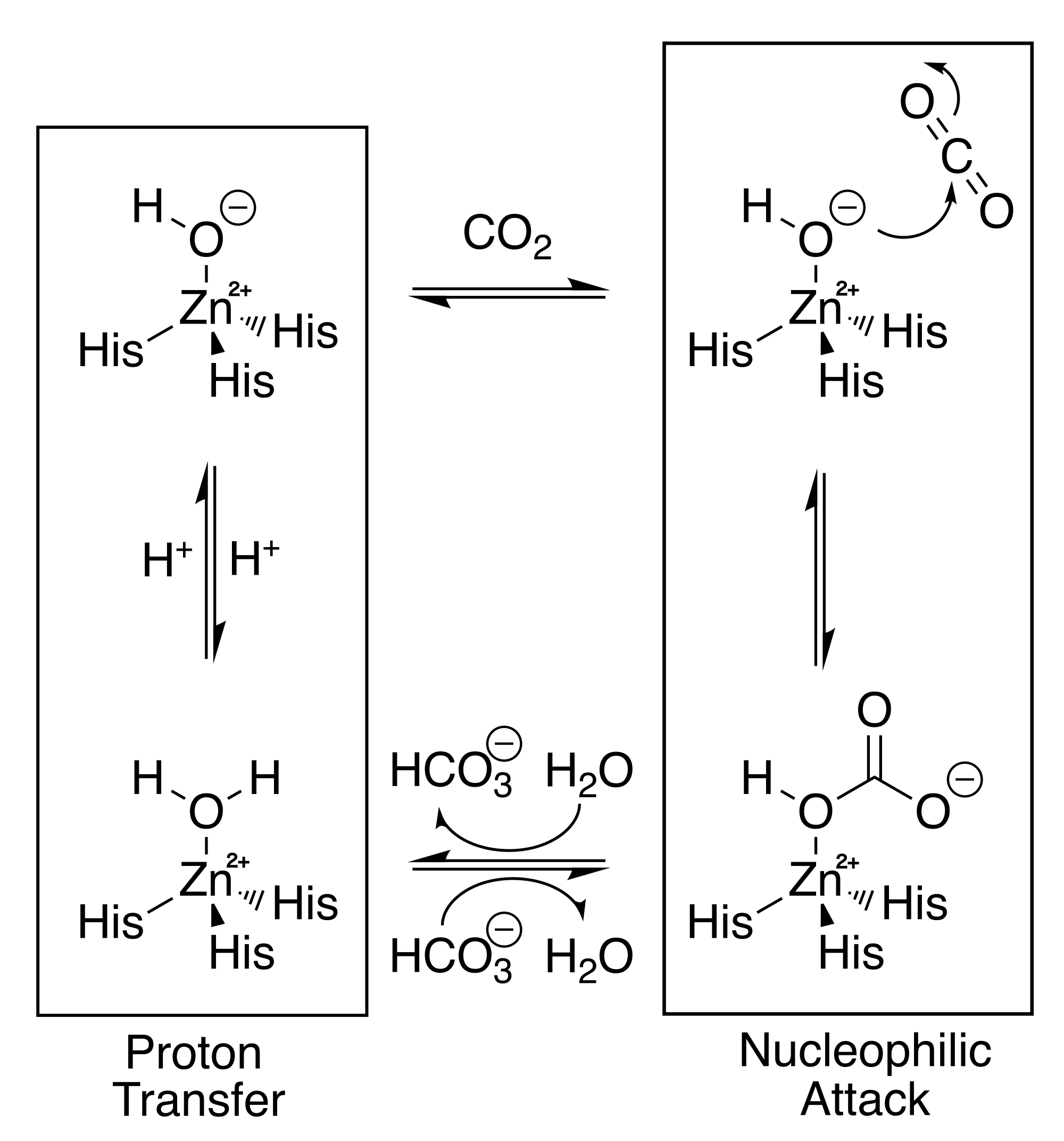}
    \caption{Catalytic cycle for the hydration of $\text{CO}_2$ in carbonic anhydrase enzymes with emphasis on the zinc ion. On the left, it is shown the proton transfer (rate-determining) step investigated in this work while the nucleophilic attack step is represented on the right.}
    \label{fig:catalytic}
\end{figure}

As envisioned by Richard Feynman quantum computing promises to simulate quantum systems efficiently\cite{Feynman1982}. The high complexity and steep scaling associated with the simulation of quantum systems on classical hardware could be alleviated with fault-tolerant quantum hardware\cite{Reiher2017,Babbush2018, Cao2019, vonBurg_2021}. This could lead to a dramatic impact on the modeling accuracy for large molecules. It is reasonable to expect that the same systems modeled at present with Kohn-Sham DFT on classical devices will, in a not-too-distant future, be treated with highly-accurate wave function methodologies on quantum computers. 
This optimistic outlook is fueled by the ongoing advancements in quantum hardware technology and the development of powerful quantum algorithms\cite{Bharti_2021,ezratty}. Despite this enormous progress, current \ac{nisq} devices are not yet fault-tolerant, and thus only capable of executing simple computational circuits.  The frequent errors on such machines will render the output meaningless if the algorithms are too complicated. Thus a range of algorithms has been developed specifically for these devices\cite{Bharti_2021}. Recently, the question was raised whether \ac{nisq} devices were able to perform useful work beyond the reach of classical devices\cite{Kim2023}, coining the expression \emph{quantum utility}. We have attempted to use a \ac{nisq} device for the task of describing the energy along the reaction coordinate of an enzymatic reaction. While these calculations are currently not beyond the reach of brute-force classical calculations, they mark a further step on path towards the realm where chemical calculations become hard to verify classically, similar to a recent study by IBM\cite{robledomoreno_2024}. It remains to note that the purpose of this study was to use modern quantum hardware to show what is currently achievable in enzymatic reaction simulation, which stands in contrast to fault tolerant algorithm developments for enzymes such as Refs.~\cite{Reiher2017, vonBurg_2021} where the purpose is to demonstrate the current state of algorithm development and what we may at some point become achievable.\\
To achieve this, we have developed and applied a fully-automated hybrid multi-level quantum computing strategy capable of modeling enzymatic reactions. This work focuses on modeling \ac{ca}, an enzyme ubiquitous in nature and first discovered in 1932 for its role in regulating the blood pH by catalyzing the hydration of carbon dioxide with the resulting release of carbonic acid\cite{Forster2000}. More recently, \ac{ca} has attracted attention globally as it offers the potential to address challenges associated with climate change, specifically when reducing the atmospheric $\text{CO}_2$ concentration\cite{Talekar2022}.\\
In particular, we have studied the energy profile of the rate-determining step of the reaction consisting of the proton transfer between a water molecule and Histidine 64 (H64), mediated by three additional solvent molecules as well as by a coordinated zinc atom\cite{pt_ref1,pt_ref2}. The full catalytic cycle for \ac{ca} is schematically shown in Fig.~\ref{fig:catalytic}.\\
A correlated wave-function method is dedicated to a part of the active site where the atoms are directly involved in the bond breaking/formation. The correlated wave function is optimized via the efficient \ac{fast} algorithm\cite{Majland_01_2023} on the IonQ Aria trapped ion quantum device. This accurately modeled portion of the system is embedded in a Kohn-Sham DFT electron density calculation in a defined QM region. The remaining system is treated classically with the \ac{gfnff}\cite{gfnff_2020,xtb_2021} method and using an ONIOM description\cite{Svensson_1996}. This is thus the first application of quantum hardware for describing an enzymatic reaction using a multi-scale embedding framework. There have been examples on hardware where less of the environment has been captured\cite{Rossmannek_01_2023} and where the multi-scale framework has not been used in a quantum hardware setting\cite{Izsak_01_2023, ma_2023, Santagati2024}. The presented framework is general and can be applied beyond biologically relevant systems, pushing quantum computing-aided molecular modeling into the realm of realistic applications.

\section{Materials and Methods}\label{sec:materials_and_methods}
Simulating the energy profile of an enzymatic reaction on a quantum computer is a complex task that necessitates a multifaceted approach. This involves several key steps which will be explained in this section: preparing the enzymatic system, identifying and optimizing the reaction coordinate, strategically dividing the molecule using a \ac{qmmm} approach, and finally coupling the classical and quantum calculations within the quantum region.

\subsection{System and reaction coordinate preparation} \label{sec:system_and_coordinate_prep}
The enzyme structure used in this work refers to the Human Carbonic Anhydrase II (PDB-ID 2CBA), Fig.~\ref{fig:2cba}, as prepared and provided in the supplementary material of Fu \emph{et al.}\cite{Fu_01_2021}.
\begin{figure}
    \centering
    \includegraphics[scale=0.20]{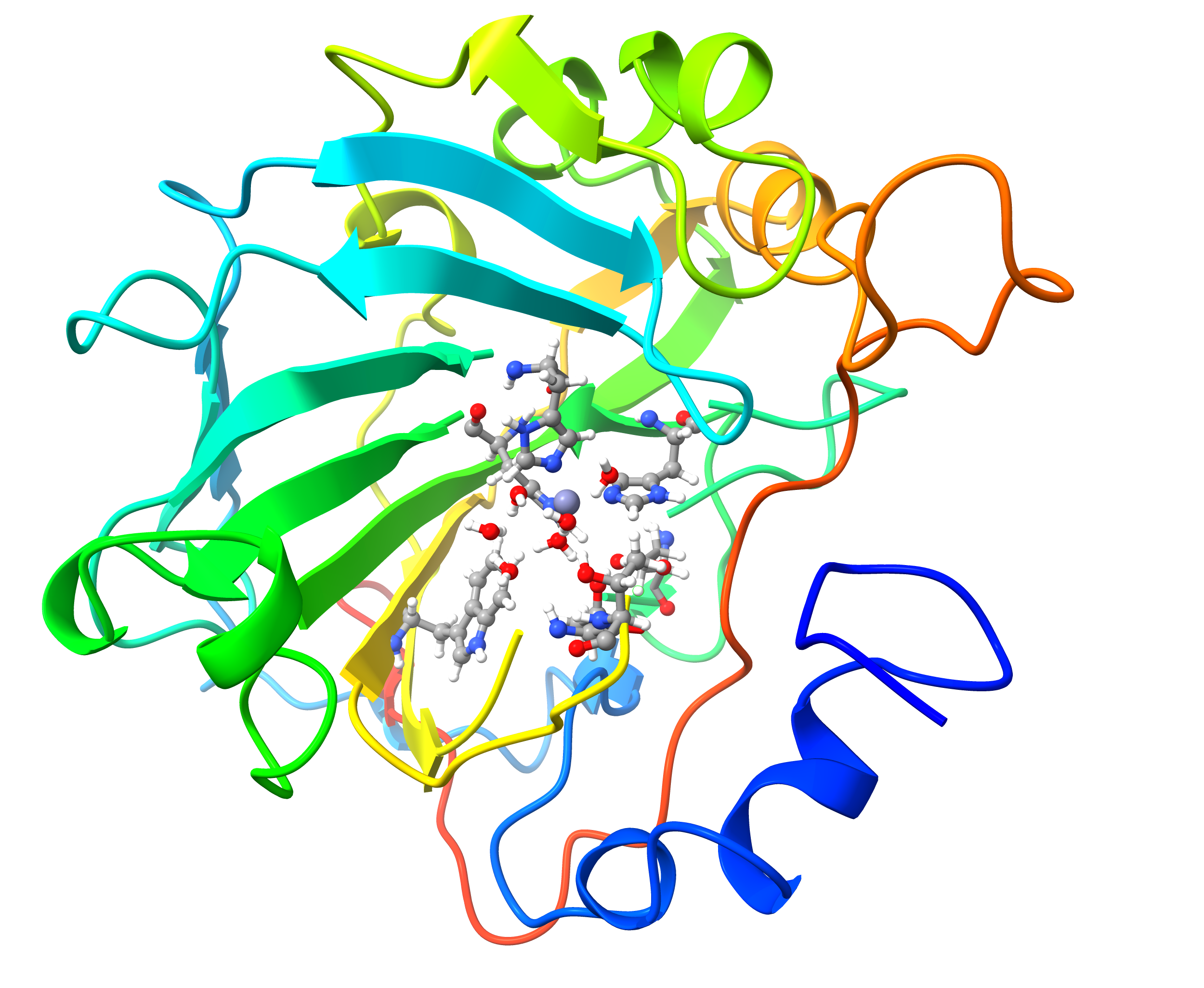}
    \caption{Representation of the Human Carbonic Anhydrase II (PDB-ID 2CBA). Explicit solvent molecules have been removed except those close to the active site. The portion of the enzyme-treated quantum mechanically is shown with a ball-and-stick representation while the rainbow ribbon representation indicates the remaining portion of the system.}
    \label{fig:2cba}
\end{figure}
The reaction coordinate was identified using our implementation of the \ac{neb} method~\cite{Jonsson_1998} while modeling the structures with the ONIOM method, using \ac{gfnff}~\cite{Spicher_01_2020} method as the low-level method and the \ac{gfn2xtb}\cite{Bannwarth_2019} method as the high-level method. Both these methods were used through the \ac{xtb}~\cite{xtb_2021} program package. 

To construct the initial path for the \ac{neb} calculation, suitable reactant and product structures first had to be obtained. Our reactant structure was found by optimizing the geometry of the reactant structure from Fu \emph{et al.}\cite{Fu_01_2021} using our chosen ONIOM[\ac{gfn2xtb}:\ac{gfnff}] model (with implicit water, using the \ac{alpb} model\cite{ehlert_robust_2021}). The geometry-optimization was converged to a maximum atomic gradient component of $\SI{e-3}{\atomicunit} \approx \SI{1.2}{\kilo\calorie\per\mole\per\angstrom}$. The product structure corresponding to the equivalent overall protein conformation was determined by manually moving only the hydrogen atoms involved in the proton transfer, to form a structure with the proper product topology; this structure was then similarly geometry optimized, and the initial path for the \ac{neb} calculation was obtained by linearly interpolating atomic positions between these reactant/product structures.

By constructing the \ac{neb} input in this way, we achieved minimal motion of the remote chemical groups of the enzyme, ensuring that there were only contributions to energy changes along the reaction path, that stemmed from conformational rearrangements associated with the proton transfer reaction. In a typical molecular dynamics simulation, this problem would be handled by adequately sampling the conformational space, which is prohibitive on a quantum computer due to the vast amount of single-point energy evaluations required. Instead, we have chosen a sample-free approach resulting in reaction enthalpies rather than free energies. This effective freezing of the rest of the protein is necessary for the correct interpretation of the energy path. It should be noted that any tunneling effects in the proton transfer reaction are neglected in this study due to the elevated computational expense for an accurate description.

The \ac{neb} path optimization was also converged to a maximum atomic gradient component of $\SI{e-3}{\atomicunit} \approx \SI{1.2}{\kilo\calorie\per\mole\per\angstrom}$. The resulting 10 points provide the input structures for the accurate correlated wave function calculations performed on quantum hardware with the multi-layer embedding strategy described below. The full enzyme structures along the reaction coordinate can be found in the supporting information material.

\subsection{Multi-layer embedding}
\begin{figure}[p!]
    \centering
    \includegraphics[scale=0.95]{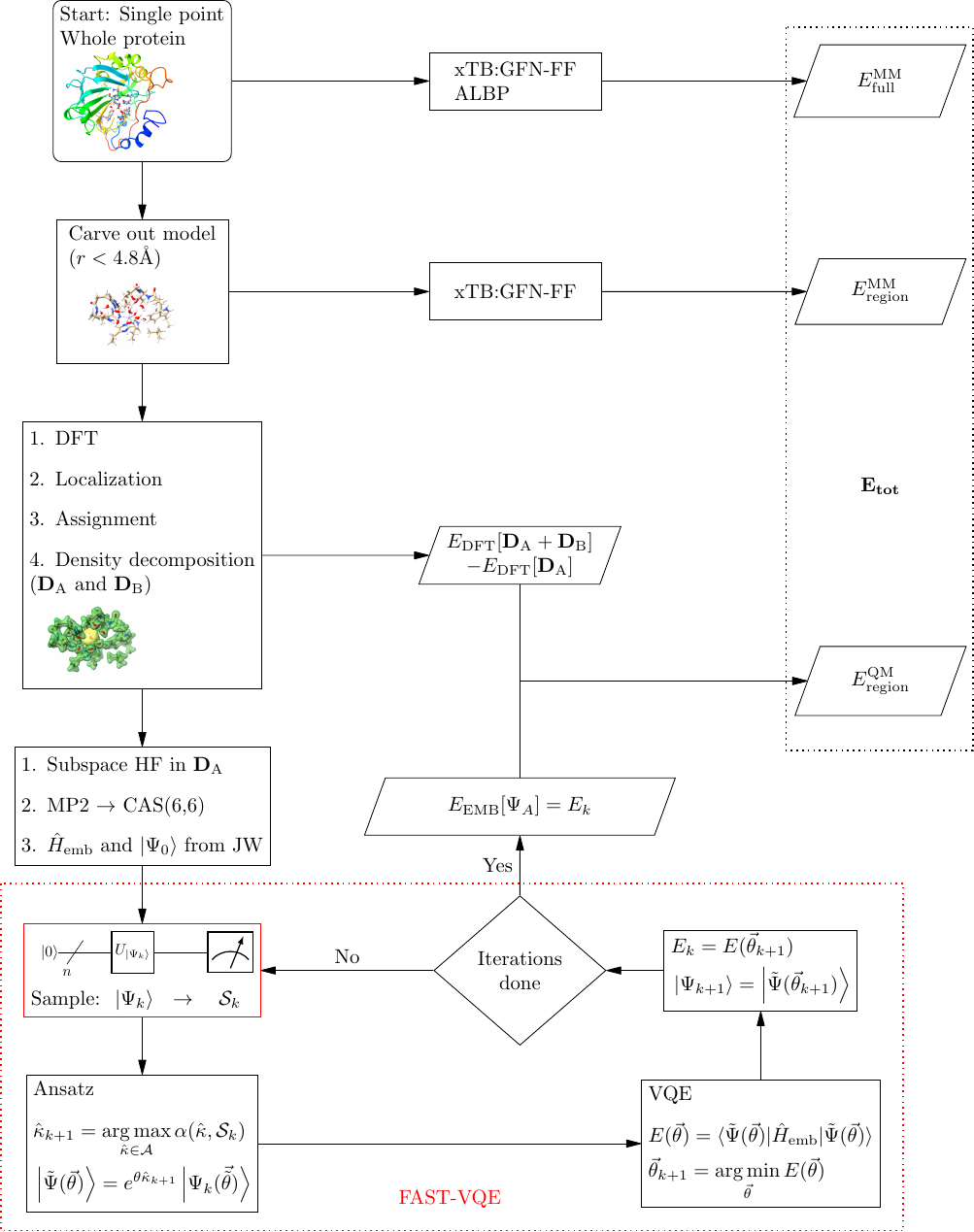}
    \caption{Schematic representation of the overall \ac{qmmm} calculation where  the intermediates required to evalate $E_\text{tot}$ of Eq.~\eqref{eq:oniom} is given in the black dotted rectangle on the right. The \ac{wfdft} calculation, is given in the lower part of the flow chart and the \ac{fast} calculation is represented in the red dotted rectangle. The wave function sampling that is executed on the quantum computer is represented by the solid red rectangle.}
    \label{fig:calc_overview}
\end{figure}

The present section describes the fully automated multi-layer embedding implementation used for this project. Besides hyper-parameter settings like the cutoff radius for the \ac{qm} region and certain algorithmic choices, no additional user-input is required. An overview over the single point multi-layer ebmedding calculation method is given in Fig.~\ref{fig:calc_overview}, this scheme is executed for each point on the reaction coodinate.\\

Multiple layers of embedding were used for the quantum calculation while adequately treating the environment. The outermost layer of the protein is coupled to the external environment by the \ac{alpb} continuum model for water implemented in the \ac{xtb}~\cite{ehlert_robust_2021, xtb_2021, Spicher_01_2020} package. The outermost portion of the protein itself is described by \ac{mm} with the \ac{gfnff} force field. This classically described region is coupled to a region about 4.8 Å in radius around the active site. The active site is described by \ac{qm} by embedding a \ac{fast} wave function into the region described by \ac{dft}. The driver for the \ac{qm} calculations on classical hardware is PySCF~\cite{Sun_01_2015, Sun_01_2018, Sun_01_2020} while the implementation of the energy component evaluated on a quantum computer is discussed profusely later in this section.\\

\begin{figure}[!htb]
    \centering
    \includegraphics[width=.7\linewidth]{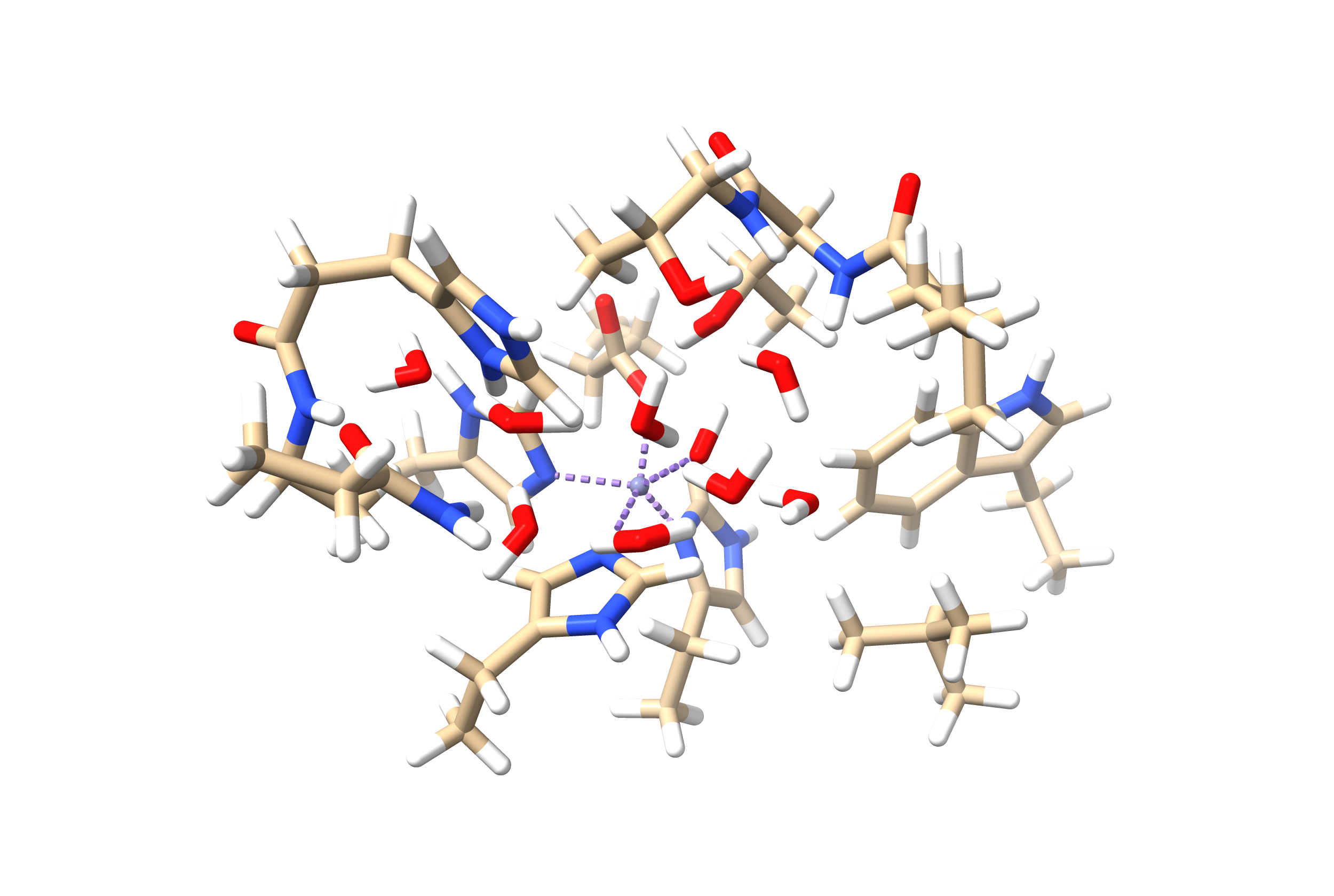}
    \caption{The portion of the enzyme defining the QM region with all residues that have an overlap with a ball of 4.8~Å radius around the active site. The model thus includes the zinc atom, the key water molecules as well and the residues involved in the proton transfer process.}
    \label{fig:qm_region_details_ball_and_stick}
\end{figure}

Since we use an ONIOM\cite{Svensson_1996} description of the system, we may write the total energy as
\begin{equation}\label{eq:oniom}
    E_{\text{tot}} = E^{\text{MM}}_\text{full} - E^\text{MM}_\text{region} + E^\text{QM}_\text{region},
\end{equation}
where the obvious approximations commonly encountered in molecular modeling may be summarized as 
\begin{enumerate}
    \item The continuum water model is only accounted for in $E^\text{MM}_\text{full}$ and neglected in the inner region calculations as the water molecules around the active site have been modeled explicitly.
    \item When partitioning the system, hydrogen caps are introduced in the region on the boundaries between the \ac{mm} and \ac{qm} regions as typically done in \ac{qmmm} methods. 
\end{enumerate}
For the \ac{qm} calculations around the active site, we include all residues and molecules that have at least one atom within a 4.8~Å radius around the active site (or more specifically, around the focus water oxygen atom and the two hydrogens involved in the proton transfer reaction) up to the first alpha carbon atoms outside of this radius see Fig.~\ref{fig:qm_region_details_ball_and_stick}. This applies for side chains and backbone portions. Non-peptide groups and molecules are included entirely. The connections of these alpha carbons are then cut and capped with hydrogen atoms following a standard procedure in \ac{qmmm} embedding calculations. The \ac{qm} region is determined once for the transition state structure and reused in the remaining calculations along the reaction coordinate to ensure consistent results.

An accurate description of the \ac{qm} region is particularly important as this region includes the most chemically relevant portion of the system taking part in the proton transfer reaction under consideration. We choose the \ac{wfdft} scheme~\cite{Manby_01_2012} as the method for modeling the \ac{qm} region due to its simple structure. In particular, the method combines correlated wave function methodologies with DFT, thus inheriting the high accuracy of the former while preserving a tractable computational cost necessary to model larger and more realistic molecular systems. Other analogous techniques could also be employed\cite{rubin_01_2016,Li_01_2022,Izsak_01_2023}, however, in connection with quantum computing applications, the chosen \ac{wfdft} approach provides notable advantages as it only relies on the energy of the correlated portion of the \ac{qm} region and is conceptually simple, although other algorithms are available\cite{Vorwerk2022}. Here, we use the measurement-efficient \ac{fast} quantum algorithm~\cite{Majland_01_2023} for the correlated region. Note that the \ac{wfdft} method has recently been used on a \ac{nisq} device in combination with the \ac{adapt} method\cite{Grimsley_01_2019} for modeling the potential energy surface of the nitrile group dissociation reaction in butyronitrile~\cite{Rossmannek_01_2023}.\\ 

The \ac{wfdft} embedding method has been described in detail before~\cite{Manby_01_2012, Rossmannek_01_2023} and thus we will focus on the parts that determine the accuracy of the overall calculation. Using this method, the \ac{qm} energy $E^\text{QM}_\text{region}$ of the total \ac{qmmm} embedding scheme from Eq.~\eqref{eq:oniom} may be written as
\begin{equation}
\label{eq:qmregion}
    E^\text{QM}_\text{region} = E_\text{EMB}[\Psi_A] + E_\text{DFT}[\mathbf{D}_A + \mathbf{D}_B] - E_\text{DFT}[\mathbf{D}_A],
\end{equation}
where $A$ denotes the orbital space of the high-accuracy calculation and $B$ is the complement of $A$. $E_\text{EMB}$ denotes the embedding energy of the high-accuracy calculation resulting in a wave function for atomic region $A$ and may be obtained from any correlated wave function. We use the \ac{fast} algorithm to provide this quantity. \\

To arrive at an effective embedding Hamiltonian that can be encoded and evaluated on a \ac{nisq} device, a couple of classical preparation steps are necessary, i.e.
orbital localization, orbital assignment, and active space selection, which we will discuss in the following sections. We will conclude the discussion by briefly describing the \ac{fast} algorithm.

\subsubsection{Orbital localization and assignment} \label{sec:localization_and_assignment}
To separate the energy in Eq.~\eqref{eq:qmregion} consistently, each occupied molecular orbital must be assigned to one of the two subsystems A and B respectively. This is done by obtaining localized molecular orbitals through a suitable unitary transformation. Several standard algorithms for this task are available and we chose intrinsic bonding orbitals\cite{ibo}. Note that \ac{wfdft} only requires localized occupied orbitals while there is no restriction on the form of the virtual orbitals. We will use this fact in the active space selection.\\

After the localization, the orbitals need to be assigned to atomic centers and several different choices are possible. We have chosen to use a simple scheme based on orbital expansion coefficients, \emph{cf.} the appendix appendix.

\subsubsection{Active space selection} \label{sec:active_space_selection}
The orbital subspace assigned to subsystem A is typically too large to encode the effective electronic Hamiltonian into a current \ac{nisq} device. To efficiently fit the problem size to the available hardware, it is necessary to have a procedure in place that constructs a consistent subspace Hamiltonian for each point across the entire PES. Therefore, we will use a frozen natural orbital (FNO) approach\cite{Sosa_1989} based on MP2 amplitudes. Note that for this work, we have chosen the active space solely based on the desire to construct a non-trivial subspace for the description of correlation effects within the available number of qubits and within a regime that may still be simulated exactly. The effect of the orbitals under the reaction were no major concern for this study. For a precise description of how the active space was constructed, please see the appendix. In a future implementation, it may be fruitful to include some recent CAS construction method\cite{Bensberg2023, Ding2023} instead.\\

\subsubsection{Subspace calculations}
After defining the active space a subspace Hamiltonian is constructed and prepared for the correlated calculation using the Jordan-Wigner mapping\cite{Jordan1928}, which was chosen because it is directly compatible with the \ac{qeb}~\cite{Yordanov_01_2021} operators chosen elsewhere in the implementation, \emph{vide infra}.
We then use the \ac{fast} algorithm~\cite{Majland_01_2023} for all correlated electronic structure calculations on a quantum computer. Here we provide the necessary background that allows us to discuss the computational details.\\

In \ac{fast}, a wave function ansatz is built adaptively by adding parameterized operators from an operator pool $\mathcal{A}$ to the ansatz in each iteration. The ansatz at the $k$th iteration of the algorithm may be written as
\begin{equation}
\label{eq:ansatz}
    \ket{\Psi_{k}} = \prod_{i=k}^1 e^{\theta_i\hat{\kappa}_i} \ket{\text{HF}},
\end{equation}
 where the excitation operator $\hat{\kappa}_i$ along with the tunable parameter $\theta_i$ are introduced at iteration $i$ and $\ket{\text{HF}}$ is the \ac{hf} reference state. Using exact excitation operators would result in too deep circuits, thus we are using  a adjusted version of the \ac{qeb}~\cite{Yordanov_01_2021} operators in the operator pool $\mathcal{A}$ containing only one and two-body particle-hole excitations, see Figs.~\ref{fig:singles-circuit} and~\ref{fig:doubles-circuit} in the appendix.
 
This adaptive approach was originally introduced by the \ac{adapt} algorithm~\cite{Grimsley_01_2019}, yet there is a difference in how the operators are selected in the \ac{fast} algorithm. In iteration $k$ of the algorithm, each operator $\hat{\kappa} \in \mathcal{A}$ is tested with an importance metric.

In \ac{adapt}, the metric $g(\hat{\kappa})$ used to assign importance to each operator in $\mathcal{A}$ relies on the energy gradient computed with respect to the associated operator parameter $\theta$.  To motivate this metric, consider the preliminary wave function 
\begin{equation}
\ket{\Psi_{k+1}} = e^{\theta \hat{\kappa}} \ket{\Psi_{k}},
\end{equation}
for all $\hat{\kappa} \in \mathcal{A}$ and where $\theta$ has not been optimized. One may then evaluate the importance of each operator by its gradient at $\theta = 0$ as given in Eq.~\eqref{eq:grad_adapt}, whose evaluate requires a complete measurement for each operator in the operator pool $\mathcal{A}$.
\begin{equation}
\label{eq:grad_adapt} 
    \begin{split}
 g(\hat{\kappa})&=\biggl(\frac{\partial E^{(k+1)}}{\partial \theta}\biggr)_{\theta =0} \\
 &=\bra{\Psi^{(k+1)}}[\hat{H},\hat{\kappa}]\ket{\Psi^{(k+1)}}.
 \end{split}
\end{equation}

Instead, in \ac{fast}, the importance of an operator for the next $k+1$ iteration is calculated by the heuristic gradient
\begin{equation}
\label{eq:heuristic}
    \alpha(\hat{\kappa}, \mathcal{S}_k)= \sum_{D_i \in \mathcal{S}_k}\sum_{D_j \in \mathcal{S}_k} \bra{D_i}\hat{\kappa} \hat{H} \ket{D_j}, \quad\forall \kappa \in \mathcal{A},
\end{equation} 
where a multi-set of Slater determinants $\mathcal{S}_k$ is sampled from $|\Psi^{(k)}\rangle$ in the computational basis (for a motivation of this metric, see Ref.~\cite{Majland_01_2023}).
In contrast to measuring the expectation values required in Eq.~\eqref{eq:grad_adapt}, sampling is rather efficient on a quantum device and thus Eq.~\eqref{eq:heuristic} may be evaluated efficiently on a classical computer. This reduces the measurement overhead per iteration to just 1 compared to the at least $O(N^6)$ measurements (N being the number of orbitals in the system) required in the \ac{adapt} method. This has proven to be a simple and robust procedure for constructing wave functions on a quantum computer.\\

Validation of \ac{fast} results is straightforward as we can use the subspace \ac{hf} energy and leverage an additional subspace \ac{casci} calculation. The former provides the starting point for \ac{fast} and the latter provides essentially exact solutions within this space, acting as benchmarks for FAST-VQE's performance. This approach is feasible because exact diagonalization with \ac{casci} in these small active spaces is computationally tractable with classical computers.

\subsection{Computational details}
\begin{figure}[!htb]
    \centering
    \includegraphics[width=.7\linewidth]{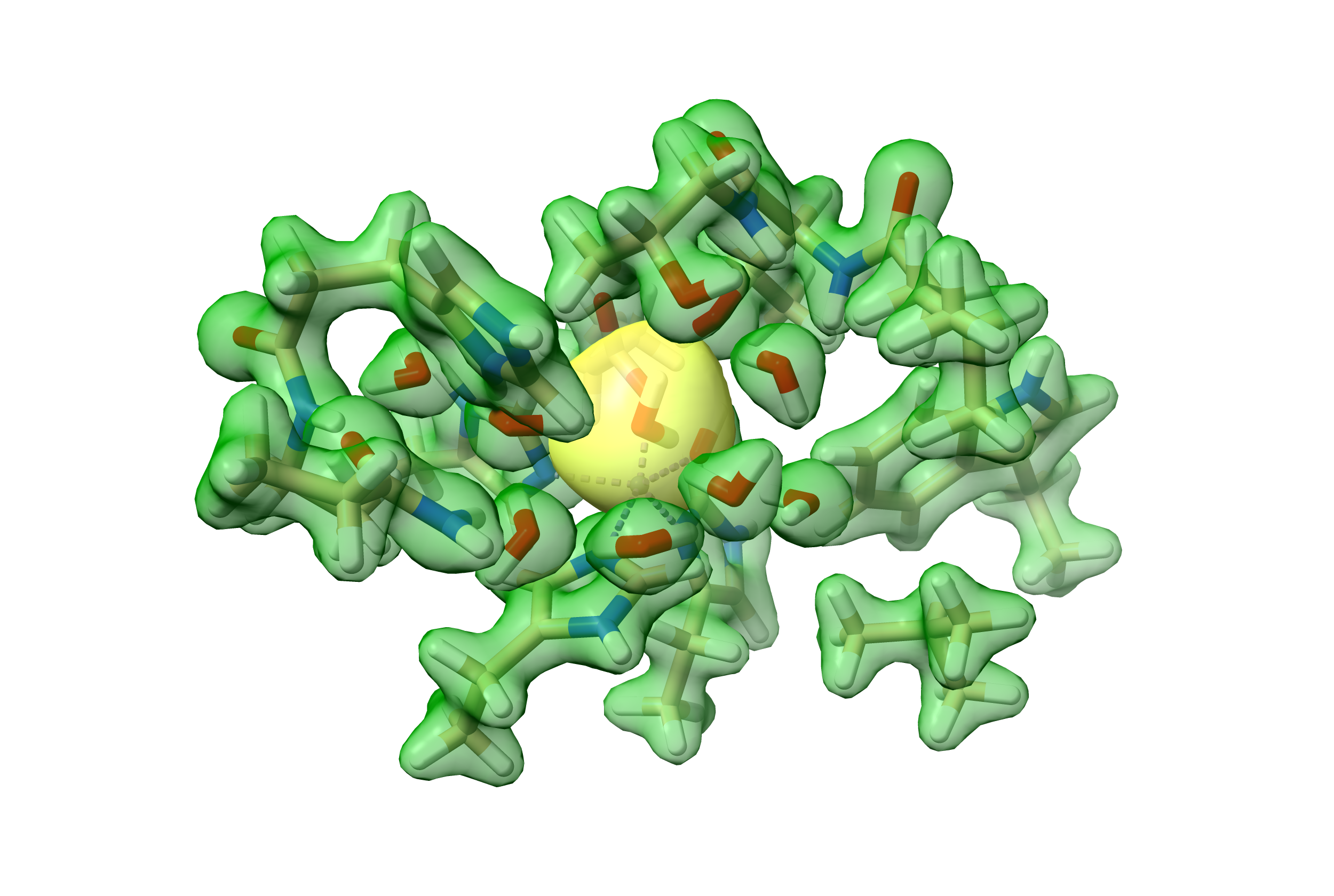}
    \caption{The model of the system from Fig.~\ref{fig:qm_region_details_ball_and_stick} including the electron densities of both fragment A (light lime green) and fragment B (darker green) arising from the density matrices $\mathbf{D}_\text{A}$ and $\mathbf{D}_\text{B}$ of the system partitioned as described earlier.}
    \label{fig:qm_region_details_density}
d\end{figure}

Fragment $A$, modeled via a correlated wave function, has been chosen to include the oxygen atom in the middle of the three water molecules in the proton transfer reaction and the two moving hydrogen atoms. This region has five local occupied orbitals assigned at the transition state geometry (point 5 on the PES) that serves as a reference. Using the procedure outlined previously, five occupied orbitals are assigned to the fragment and an active space is calculated for each point. The \ac{qm} region is shown in Fig.~\ref{fig:qm_region_details_ball_and_stick}, while Fig.~\ref{fig:qm_region_details_density} also shows the electron density of fragments A and B respectively.\\
As discussed, the correlated wave function is generated from a complete active space of 6 electrons in 6 orbitals, embedded into an electron density computed with the B3LYP\cite{Stephens1994, Becke1997} functional while the basis-set employed is Jensen's pcseg-1\cite{pcseg}.

All SCF parameters were precomputed on Amazon EC2 \texttt{c5.12xlarge} instances, with 48 Intel Xeon Platinum 8000 series processors and 96Gb memory, and stored into suitable files. These files were then used to restart the calculations using Amazon Braket Hybrid Jobs on IonQ's Aria 1 quantum device. Parameters were optimized using an embedded state vector simulator. Additionally, three single points (initial, transition state, and product structures) were computed on Rigetti's Aspen-M-3 superconducting quantum device allowing us to compare different quantum hardware technologies as discussed later in the manuscript. For all calculations, we chose to extend the ansatz in Eq.~\eqref{eq:ansatz} with one operator in each iteration and all experiments were conducted using 1024 shots on Aria 1 and Aspen-M-3.

\subsection{Circuit optimization} \label{sec:circuit_optimization}
Naively constructing the quantum circuits from the parametrized ansatz typically leads to long circuits that may not be executable on the targeted quantum hardware. On Amazon Braket, IonQ's Aria has a limit of 950 single qubit gates which we encountered beyond 40 iterations of \ac{fast}.
To reduce overall error and tackle hardware limitations, we apply circuit optimization.
We achieve an effective gate reduction we combine two existing techniques for reducing the number of CNOT gates~\cite{shaik2024} and one-qubit gates~\cite{sivarajah2020t}.

For the CNOT gate reduction, global optimization is not practical, thus peephole techniques are promising where slices of the circuits are optimized individually.
We applied a standard slice-and-optimize strategy for optimizing CNOT sub-circuits.
While several CNOT synthesis techniques exist\cite{Iwama_2002, Meuli_2018,jiang2022,Meijer_van_de_Griend_2023, Gheorghiu_2023}, we applied the SAT-based synthesis for optimal CNOT circuits by Shaik \emph{et al}~\cite{shaik2024}.
For one-qubit gate reduction, we employ standard Clifford Simplification from TKET~\cite{sivarajah2020t}, which are expected to perform well for circuits built from \ac{qeb} operators as they mainly contain Clifford gates.

\section{Results and Discussion}
In this section, we will focus on the description of the results, starting with a classically identified reaction coordinate, from which ten points are chosen for the evaluation on a quantum computer using the multi-scale embedding framework, followed by reporting the discussion of our results obtained on various quantum hardware.

\subsection{Quantum experiments}
\begin{figure}[!htb]
\centering
   \includegraphics[width=0.6\linewidth]{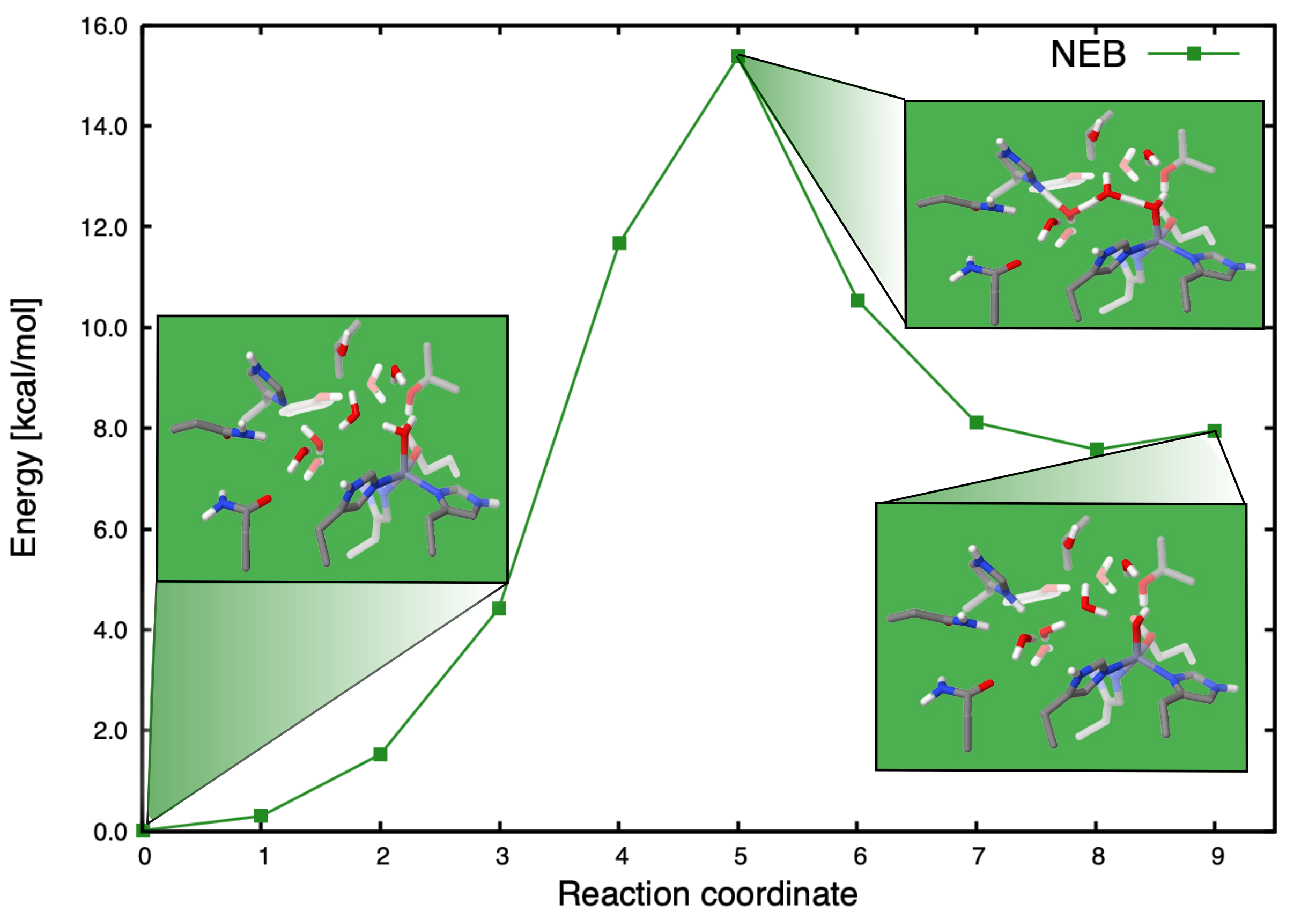}
   \caption{Potential energy surface obtained via a \ac{neb} calculation at the GFN2-xTB/GFN-FF level. The core portion of the initial, transition state and product structures used in the reaction path optimization are shown along the potential energy surface. An animation of the reaction can be found in the supporting information. The minimum at the end of the path is shallow and, with the convergence settings used, NEB was able to identify a point with a lower energy (point 8) than the optimized product structure (point 9).}
   \label{fig:neb-pes} 
\end{figure}

The reaction coordinate, identified as described in the Materials and Methods section, displays an energy barrier of circa 15.5 kcal/mol with respect to the initial reactants, slightly higher that the 10.5 kcal/mol barrier computed by Fu and coworkers, albeit their result refers to the free energy of the reaction and thus includes entropic contributions~\cite{Fu_01_2021}. The resulting PES is shown in Fig.~\ref{fig:neb-pes}. Note that because the product region minimum is broad and shallow compared to the gradient convergence threshold of $\SI{e-3}{\atomicunit} \approx \SI{1.2}{\kilo\calorie\per\mole\per\angstrom}$ the \ac{neb} calculation has found another product structure that is slightly lower in energy than the path endpoint. 

\begin{figure}[!htb]
\centering
   \includegraphics[width=0.6\linewidth]{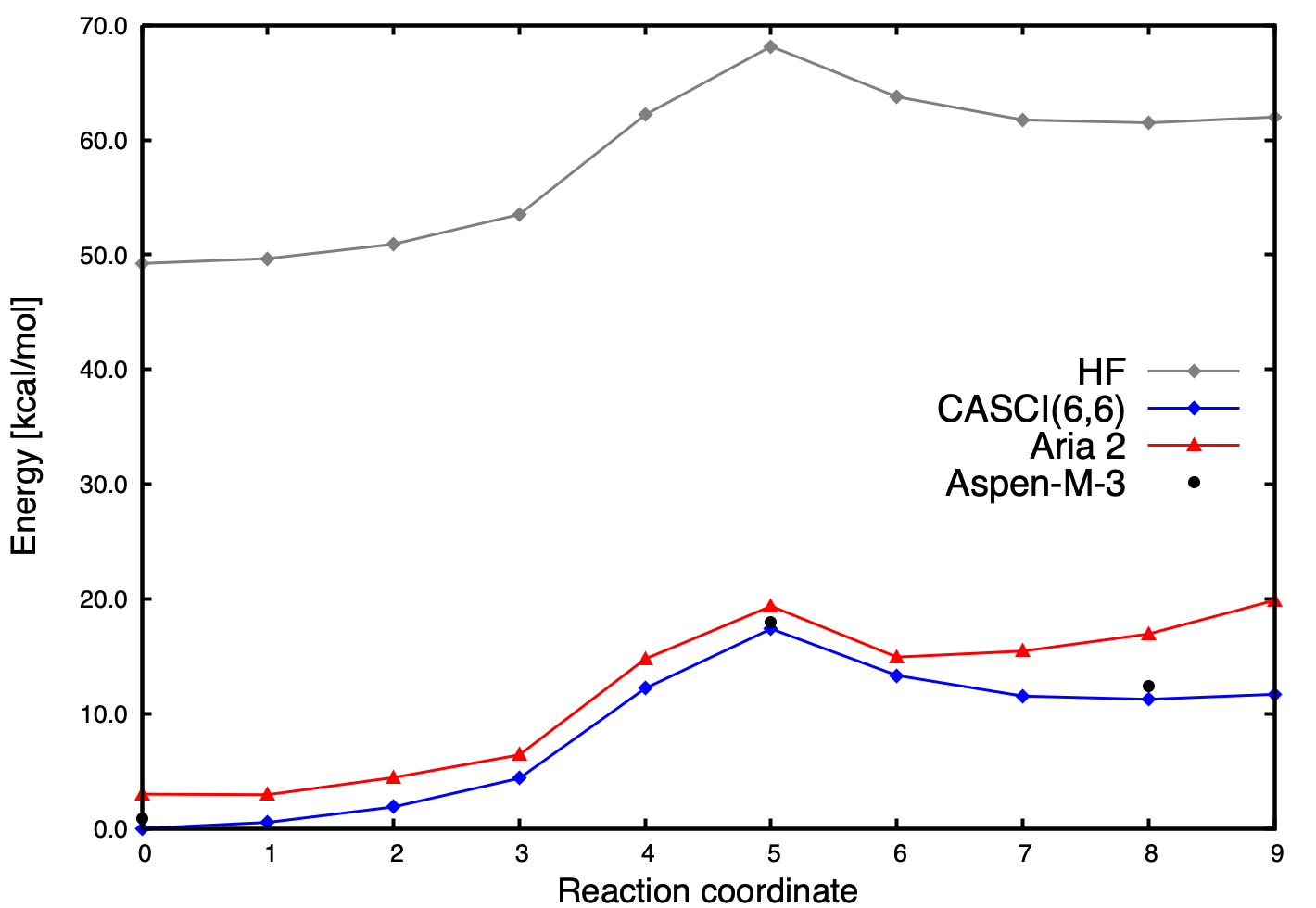}
   \caption{Potential energy surfaces obtained on classical hardware (grey and blue) as well as on IonQ's Aria 2 (red) and Rigetti Aspen-M-3 quantum hardware (black), the latter will be discussed later in the manuscript. The grey (blue) curve refers to a \ac{hf} (\ac{casci}) treatment of the embedded subsystem A providing the starting point (exact reference) for a \ac{fast} calculation. All curves have been shifted such that the lowest energy point on the CAS(6,6) curve has zero energy.}
   \label{fig:pesses} 
\end{figure}

The reaction path obtained provides the 10 input structures for the \ac{wfdft} correlated calculations. Fig.~\ref{fig:pesses} shows the PES obtained with the \ac{wfdft} method on classical hardware providing the benchmark, called CAS(6,6), and the results obtained on the Aria-1 IonQ quantum system via the \ac{fast} algorithm with 40 iterations.
Since the used methods are all variational, the PES representing the classically exact CAS(6,6) benchmark (blue curve) is always below the \ac{fast} solution (red curve), which in turn is always below the \ac{hf} solution (black curve), corresponding to ordering the methods by the number of free parameters. In general, the \ac{fast} solution is very close to the exact CAS(6,6) curve, although the last three points on the PES exhibit a larger error.

\begin{figure}
    \centering
    \includegraphics[scale=0.2]{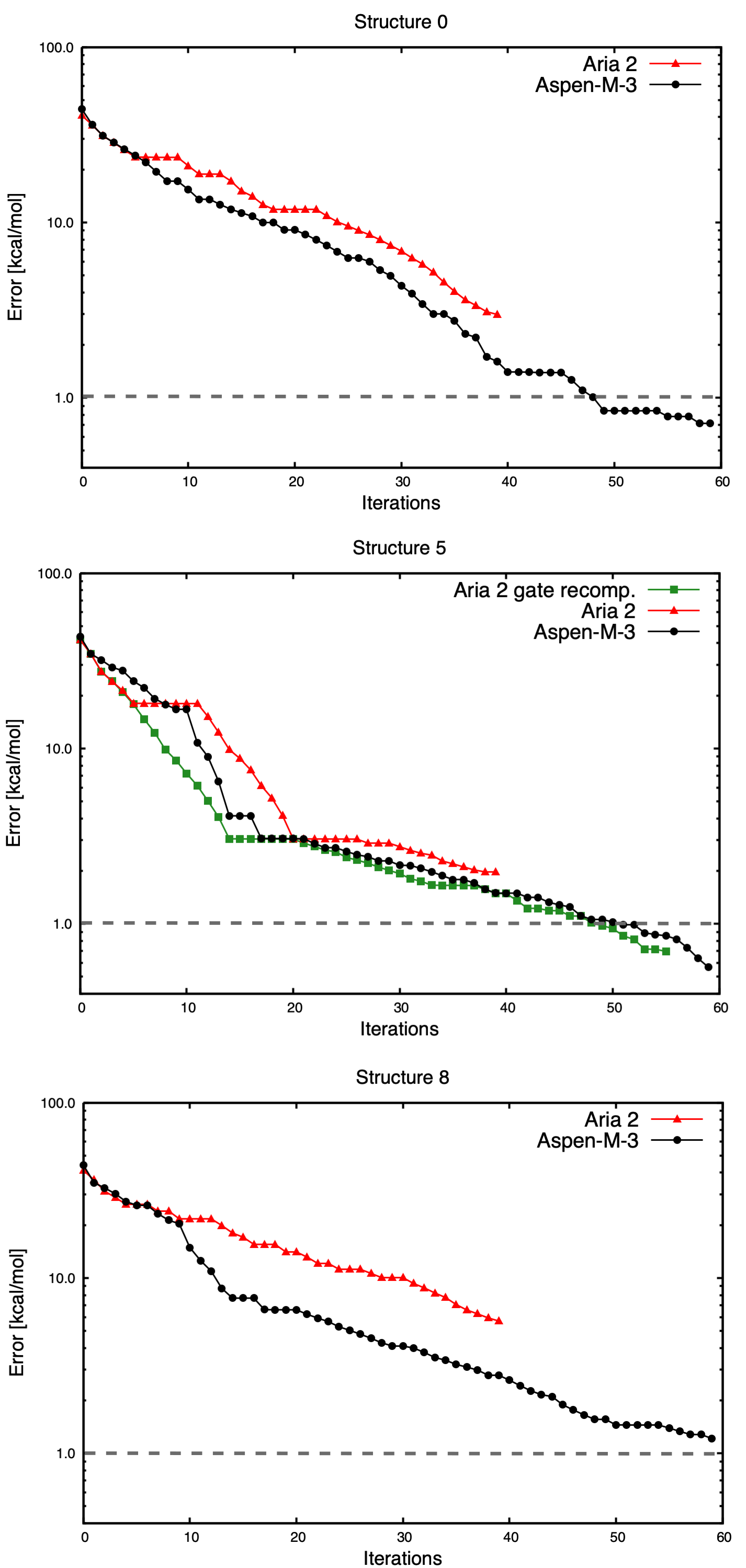}
	\caption{The subfigures show the error (in kcal/mol) between the \ac{fast} solution and the classical computing benchmark (CAS(6,6))as a function of the number of iterations computed with Aria (IonQ) and Aspen-M-3 (Rigetti) quantum systems. The plots refer to structures 0, 5, and 8 along the proton transfer reaction coordinate. Chemical accuracy defined as 1~kcal/mol is highlighted in all plots by a dashed line. For structure 5, we additionally show the convergence pattern for Aria using the gate recompilation technique, described in the Circuit optimization section, up to 55 iterations. A logarithmic scale is used for the y-axes.}
    \label{fig:error_per_point}
\end{figure}

The choice of the active space was mainly dictated by the number of qubits available for the calculation, namely 12 that corresponds to 12 spin-orbitals thus 6 spatial ones included in the CAS construction where these were chosen based on MP2 natural orbital with occupation numbers can be rounded to 1. This choice of orbitals allows for capturing the largest part of correlation effects. We also note, however, that the electronic structure of the system included in the core QC region, is not very complicated as the orbitals are \emph{de facto} perturbed by the two electron-free protons moving along the reaction coordinate.\\
Although being efficient when building an ansatz adaptively, an increased number of \ac{fast} iterations comes with deeper circuits which may exceed the specific quantum hardware's capabilities. For the Aria-1 quantum device on Amazon Braket, a maximum of 950 1 qubit gates are available, limiting the total number of \ac{fast} iterations to 40 with a standard gate compilation protocol. This impacts the overall accuracy of the calculation since it implies fewer free parameters to model the wave function.  Convergence plots are displayed in Fig.~\ref{fig:error_per_point} showing how the deviation from the classical solution evolves as a function of the iterations for three selected points including initial-, transition- and final-state structures along the PES. The convergence plots of the entire PES are given in the Supplementary Material. \\

To further push and probe convergence in terms of iterations, we selected a reactant, transition state, and product structures, structures 0,5 and 8, respectively, and ran the calculations with the \ac{fast} algorithm on Rigetti's Aspen-M-3 system, powered by superconducting qubit-based quantum processors. There was no restriction on the number of gates and this device allowed us to reach 60 FAST-VQE iterations, exhibiting a continued consistent convergence towards the CAS(6,6) results in the additional 20 iterations, see Fig.~\ref{fig:error_per_point}. Despite the difference in the number of iterations, the convergence on IonQ's Aria, and Rigetti's Aspen-M-3 results show similar convergence patterns.

We note that with 60 iterations, it is possible to obtain errors very close or even below chemical accuracy (1 kcal/mol) along the whole reaction coordinate as exemplified in the three points above. This underlines how current advanced quantum hardware combined with state-of-the-art quantum algorithms such as the \ac{fast} approach can reach results inconceivable until recently in terms of accuracy and applicability.

Using the circuit optimization techniques described previously, it was possible to reach 55 \ac{fast} iterations due to a reduction of the ansatz circuits from 1016 to 660 single qubit gates and 540 to 533 two-qubit gates on IonQ's Aria 1 system. The calculation exhibits the same convergence pattern as Rigetti's Aspen-M-3 system for the last 15 iterations and thus reaches chemical accuracy for the transition state structure (structure 5), see the central plot in Fig.~\ref{fig:error_per_point}.

\section{Conclusions}
We have developed a fully automatic multi-scale quantum computational framework and applied it for the $\text{CO}_2$ hydration catalysed by Carbonic Anhydrase running on actual quantum hardware. To our knowledge, this is the first quantum computing application in modeling enzymatic reactions. The core region of the enzyme was treated quantum mechanically and partitioned into a higher accuracy part modeled with a correlated wave function optimized on quantum hardware via the efficient \ac{fast} algorithm. This accurate region was embedded into the remaining portion of the system modeled with \ac{dft} and the results show excellent agreement with classical computing benchmark calculations. In particular, we demonstrated that chemical accuracy may be reached consistently even on today's error-prone quantum devices, provided that the depth of the quantum circuits can be kept below imposed gate limitations. Effective circuit optimization strategies help achieve shallower circuits. Eventually, more resilient or fault-tolerant architectures must take over to allow high-accuracy calculations when encountering complicated electronic structures. This problem is exacerbated when qubits are not all-to-all connected and more qubits are used necessitating additional operations and increasing the circuit depths. With respect to quantum utility, the remaining bottleneck in our framework is the parameter optimization via the \ac{vqe} algorithm which is currently run using a state vector simulator due to its large measurement overhead. While this hampers the scalability of the presented approach, there exist techniques to avoid \ac{vqe}\cite{kanno_2023,nakagawa_2023,robledomoreno_2024}. To demonstrate quantum utility for enzymatic reaction simulations on quantum hardware, either an algorithmic change, to avoid \ac{vqe}, or higher accuracy hardware, to run \ac{vqe}, is required. \\

The proposed framework presents the first scalable implementation for routinely running enzymatic catalysis applications on a quantum computer, while achieving chemical accuracy by reducing the circuit depth using an exact gate reduction strategy. Due to a flexible implementation of the multilayer approach, it is possible to update the method as the hardware matures towards fault-tolerant machines enabling future-proof hybrid computing, e.g. by replacing \ac{fast} by a quantum phase estimation. Furthermore, the \ac{fast} algorithm itself may still be used in a fault-tolerant future as a state-preparation algorithm to improve the probability of success when running quantum phase estimation protocols. \\
Quantum phase estimation for quantum chemistry applications may be linear scaling in depth and width of the required circuits with respect to the number of orbitals and particles in the system\cite{Babbush2018,vonBurg_2021}. This enables using large active spaces, only limited by execution time and the number of available qubits, to systematically approximate the exact full-space solution. Quantum algorithms will thus eventually replace DFT when describing crucial portions of proteins or other chemical systems, finally surpassing DFT's transferability issues that hamper quantitative modeling of chemical reactions. On the other hand, for thermodynamic properties that require efficient sampling of the configuration space with millions or even billions of energy evaluations, it is difficult to foresee a path where the work is offloaded to a quantum computer. Even the best-case estimates for running quantum phase estimation on future quantum computers are about a day per energy evaluation\cite{vonBurg_2021}. Thus tremendous progress will be necessary to replace existing techniques, such as force-field-based molecular dynamics. Thus, quantum computing may be viewed as a tool that provides accuracy rather than speed in chemical calculations.\\

Nevertheless, quantum computing-based computational chemistry has a tremendous potential impact in accelerating enzyme development by complementing experimental approaches with highly predictive modeling, guiding the design and optimization of enzymes for various applications in biotechnology, medicine, and industry. Currently, activation energy calculations for enzymatic reactions are very time-consuming due to the size of the molecular system, and they may be imprecise since a multiconfigurational approach is out of reach. With the promise to tremendously speed up such precise simulations, quantum computing may lead to the development of computational tools that fit into fast-paced laboratory workflows and open up completely new possibilities. In particular, the design-build-test-learn cycle would be better informed in each round, enabling the design of new and better enzyme variants.\\

\section*{Supporting information}
We provide
\begin{enumerate}
    \item A file with the description of the accompanying xyz files (model\_description.xyz).
    \item All full-protein structures of the reaction path as concatenated xyz file (trajectory.xyz)
    \item All models of the active site and relevant neighboring residues used for the \ac{wfdft} calculations along the reaction path as concatenated xyz file (qm\_models.xyz).
    \item An animation of the proton transfer reacton for illustration purposes (model\_qm.gif).
\end{enumerate}

\section*{Data availablility}
All relevant data is available online as supporting information to this article.

\section*{Code availablility}
The \ac{mm} calculations have been carried out using the \ac{xtb}~\cite{xtb_2021} program package. Electronic structure calculations for the embedding scheme, including orbital localization, were carried out using PySCF~\cite{Sun_01_2015, Sun_01_2018, Sun_01_2020} and all quantum circuits have been implemented in Qiskit~\cite{qiskit2024}.

\section*{Acknowledgements}
We thank Amazon Web Services (AWS) for facilitating remote access to IonQ's Aria-1 system and to Rigetti's Aspen-M-3 device through Amazon Braket and supporting this project via the \href{https://aws.amazon.com/government-education/research-and-technical-computing/cloud-credit-for-research/}{AWS Cloud Credit for Research} program. This work was funded by the European Innovation Council through Accelerator grant no. 190124924.

\bibliography{main}

\newpage
\appendix
\section{Orbital assignment} \label{ch:appendix_orbital_assignment}
After some investigation, we finally opted for a simple ranking strategy for selecting the orbitals to be included in the fragment $A$ which is defined as a collection of atomic indices. Such a strategy may be constructed by ranking occupied local orbitals according to the simple weight defined as 
\begin{equation}
\label{eq:weight}
    w_i = \sum_{a \in A} \sum_{\mu \in \mathcal{B}_a} |C_{\mu i}|,
\end{equation}
where $i$ refers to the specific localized (occupied) orbital, $a$ is the index for an atom included in subsystem $A$, $\mathcal{B}_a$ is the list of atomic orbital indices associated with atom $a$, and $C_{\mu i}$ is the coefficient for the $\mu$-th atomic orbital contributing to the specific localized orbital $i$. The set $\{w_i\}$ is computed at each point along the reaction coordinate and a fixed number of orbitals is assigned to region $A$ at each point. The number of orbitals in region $A$ is somewhat arbitrary and we have therefore chosen to determine a number $n$ at the transition state structure based on 
\begin{align}
    f(i) &= \argmax_{a \in A \cup B} \sum_{\mu \in \mathcal{B}_a} |C_{\mu i}|,\\
    n &= \sum_{b \in A} \sum_{i} \delta_{b,f(i)}.
\end{align}
Here $f(i)$ is a function that maps an orbital $i$ to the atom $a$ at which $i$ has the largest cumulative weight and $n$ then counts the number of orbitals assigned to any atoms in $A$.

\section{MP2 based FNO construction}\label{ch:appendix_fno_construction}
In the occupied subspace of fragment $A$ and \emph{all} virtual orbitals a \ac{hf} calculation is required as a basis for the correlation energy calculation on a quantum device, \emph{cf.} Fig.~\ref{fig:calc_overview}. Instead of directly constructing the subspace Hamiltonian from the \ac{hf} orbitals, we perform an MP2 calculation in the subspace of the requested number of occupied orbitals, in this case 6 spin orbitals, taken from right below the Fermi-level and the full virtual space. We then use the resulting MP2 amplitudes $t_{ij}^{ab}$ to construct the virtual correlation density matrix defined in Eq.\eqref{mp2dnx}
\begin{equation}
\label{mp2dnx}
    d_{ab} = \sum_{cij} t_{ij}^{ac}t_{ij}^{cb}
\end{equation}
which may be diagonalized to define a new set of (natural) virtual orbitals which are then included according to the absolute size of the corresponding eigenvalue, starting with the largest values until the desired amount of orbitals is obtained. While not perfect, the selected correlation spaces are reasonable due to both a locality and energy/eigenvalue selection criterion and can capture some static and dynamic electronic correlation effects present in the system. Note that the active space constructed in this way was solely a choice driven by the desired to construct a non-trivial subspace for running the correlated calculation irrespective of its meaning for the chemical process. The actual size of the active space was purely defined by the number of available qubits, thus by the hardware. For the present calculations we have used 6 electrons in 6 spatial orbitals corresponding to 12 qubits (one for each spin-orbital).\\

\section{Singles and doubles circuits}\label{ch:appendix_corrected_circuits}
Here we present the singles and doubles exciation operators based on Yordanov \emph{et al}\cite{Yordanov_01_2021} as implemented for this project.
\begin{figure}
    \centering
    \scalebox{0.75}{
    \Qcircuit @C=1.0em @R=0.2em @!R { \\
                \nghost{{q}_{i} :  } & \lstick{{q}_{i} :  } & \gate{\mathrm{R_Z}\,(\mathrm{\frac{-\pi}{2}})} & \gate{\mathrm{R_X}\,(\mathrm{\frac{\pi}{2}})} & \ctrl{1} & \gate{\mathrm{R_X}\,(\mathrm{\theta})} & \ctrl{1} & \gate{\mathrm{R_X}\,(\mathrm{\frac{-\pi}{2}})} & \gate{\mathrm{R_Z}\,(\mathrm{\frac{\pi}{2}})} & \qw \\
                \nghost{{q}_{j} :  } & \lstick{{q}_{j} :  } & \gate{\mathrm{R_X}\,(\mathrm{\frac{\pi}{2}})} & \qw & \targ & \gate{\mathrm{R_Z}\,(\mathrm{\theta})} & \targ & \gate{\mathrm{R_X}\,(\mathrm{\frac{-\pi}{2}})} & \qw & \qw\\
\\ }}
    \caption{Parametrized singles excitations circuit}
    \label{fig:singles-circuit}
\end{figure}

\begin{sidewaysfigure}
    \centering
\scalebox{0.75}{
\Qcircuit @C=1.0em @R=0.2em @!R { \\
                \nghost{{q}_{i} :  } & \lstick{{q}_{i} :  } & \ctrl{1} & \ctrl{2} & \qw & \gate{\mathrm{R_Y}\,(\mathrm{\frac{\theta}{4}})} & \ctrl{1} & \gate{\mathrm{R_Y}\,(\mathrm{-\frac{\theta}{4}})} & \ctrl{3} & \gate{\mathrm{R_Y}\,(\mathrm{\frac{\theta}{4}})} & \ctrl{1} & \gate{\mathrm{R_Y}\,(\mathrm{-\frac{\theta}{4}})} & \ctrl{2} & \gate{\mathrm{R_Y}\,(\mathrm{\frac{\theta}{4}})} & \ctrl{1} & \gate{\mathrm{R_Y}\,(\mathrm{-\frac{\theta}{4}})} & \ctrl{3} & \gate{\mathrm{R_Y}\,(\mathrm{\frac{\theta}{4}})} & \ctrl{1} & \gate{\mathrm{R_Y}\,(\mathrm{-\frac{\theta}{4}})} & \ctrl{2} & \gate{\mathrm{P}\,(\mathrm{\frac{\pi}{2}})}& \qw & \ctrl{1} & \qw\\
                \nghost{{q}_{j} :  } & \lstick{{q}_{j} :  } & \targ & \qw & \gate{\mathrm{X}} & \gate{\mathrm{H}} & \targ & \qw & \qw & \qw & \targ & \qw & \qw & \qw & \targ & \qw & \qw & \qw & \targ & \gate{\mathrm{H}} & \qw & \qw & \gate{\mathrm{X}} & \targ & \qw\\
                \nghost{{q}_{k} :  } & \lstick{{q}_{k} :  } & \ctrl{1} & \targ & \qw & \qw & \qw & \qw & \qw & \qw & \qw & \gate{\mathrm{H}} & \targ & \gate{\mathrm{H}} & \qw & \qw & \qw & \qw & \gate{\mathrm{R_Y}\,(\mathrm{\frac{\pi}{2}})} & \gate{\mathrm{P}\,(\mathrm{\frac{\pi}{2}})} & \targ & \gate{\mathrm{P}\,(\mathrm{\frac{-\pi}{2}})} & \gate{\mathrm{R_Y}\,(\mathrm{\frac{-\pi}{2}})} & \ctrl{1} & \qw\\
                \nghost{{q}_{l} :  } & \lstick{{q}_{l} :  } & \targ & \qw & \gate{\mathrm{X}} & \qw & \qw & \gate{\mathrm{H}} & \targ & \qw & \qw & \qw & \qw & \qw & \qw & \qw & \targ & \gate{\mathrm{H}} & \qw & \qw & \qw & \qw & \gate{\mathrm{X}} & \targ & \qw\\
\\ }}
    \caption{Parametrized doubles exciation circuit.}
    \label{fig:doubles-circuit}
\end{sidewaysfigure}

\newpage
\section*{Graphical TOC Entry}
\begin{figure}[!htb]
\centering
\end{figure}
\end{document}